\documentstyle[nato,numreferences]{crckapb}

\begin{opening}
\title{ THREE-DIMENSIONAL COMBUSTION  IN TYPE Ia SUPERNOVAE \protect\\}

\author{A. M. Khokhlov}
\author{E. S. Oran}
\author{J. C. Wheeler}
\institute{Department of Astronomy, UT Austin, TX 78712\\ 
           Naval Research Laboratory, Washington, DC 20375}

\end{opening}

\runningtitle{COMBUSTION IN TYPE IA}

\begin{document}

\begin{abstract}

Turbulent combustion is three-dimensional. 
Turbulence in a Type~Ia supernova is driven on large scales
by the buoyancy of burning products.
The turbulent cascade penetrates down to very
small scales, and makes the rate of deflagration
independent of the microphysics. 
The competition between the turbulent cascade and
the freeze-out of turbulent motions due to stellar expansion
determines the largest scale participating in the cascade.
This sets the bulk rate of a deflagration in a supernova. 
The freeze-out 
limits the bulk rate of deflagration to a value that makes 
a powerful explosion impossible.
Two-dimensional simulations cannot capture these 
essential elements of
turbulent combustion, even in principle.

A powerful delayed detonation explosion can take place 
if the burning makes a transition to a detonation.
A deflagration to detonation transition (DDT) 
can occur in a layer of
mixed cold fuel and hot burning products created 
either inside an active
turbulent burning region by a high intensity 
of turbulence, or by mixing cold fuel with ashes 
of a dead deflagration
front during the global pulsation of a star, or by both.
\end{abstract}

\section{Introduction}

\medskip\noindent
We discuss three major topics: 

\smallskip\noindent
 1. {\sl Turbulent combustion in a uniform gravitational field}.
We summarize  the results of three-dimensional numerical
modeling of buoyancy-driven turbulent burning in supernovae 
and in terrestrial environments
and point out the possibility of terrestrial experiments 
designed specifically for modeling terrestrial turbulent
combustion and understanding combustion in supernovae. 

\smallskip\noindent
2. {\sl Turbulent combustion in an expanding star}.  
We describe the {\sl first} three-dimensional
fluid dynamics simulations of the deflagration of 
an entire carbon-oxygen
white dwarf exploding as a Type~Ia supernova. 
The results clearly point to the
important role of the expansion freeze-out of turbulent motions. 

\smallskip\noindent
3. {\sl Deflagration to detonation transition (DDT)}.
Based on a
theory of deflagration to detonation transition (DDT)  
in an unconfined environment, and on the estimates of 
the turbulent velocity obtained from the
three-dimensional simulations, we analyze the possibility of the 
DDT in supernovae, and its relevance to the 
(pulsating) delayed detonation
models of Type~Ia supernova events.

\smallskip\noindent
Limitations of the current state of numerical modeling,
and important problems for the future 
are discussed in the conclusion.

\medskip
\centerline{\bf Nomenclature}

\begin{table}[htb]
\begin{center}
\begin{tabular}{llcll}
$R_*$ &-- stellar radius & & $U_e$ &-- expansion velocity \\
$g$   &-- gravitational acceleration & & $R_f$ &-- flame radius \\
$L_f$   &-- freeze-out length & & $\tau_e$ &-- expansion time-scale \\
$L$   &-- column width & & $S_l$ &-- laminar flame speed \\
$\delta_l$ &-- laminar flame thickness & & ${\rm F}$ &-- Froude number \\
$U'$ &-- turbulent velocity & & $\lambda_G$ &-- Gibson scale \\
$\lambda_c$ &-- critical wavelength for & & $\alpha>1$ &-- density ratio across \\
&~~the onset of turbulent burning  & & &~~laminar flame front \\

$D$   &-- detonation wave speed & &  $\delta_d$ &-- detonation wave thickness \\
$\lambda_{d,c}$ &-- detonation cell size & & $\lambda_k$ &-- detonation kernel size\\
$L_i$ &-- detonation initiation length & & $U_i$ &-- detonation initiation velocity \\

\end{tabular}
\end{center}
\end{table}

\section{Turbulent combustion in a uniform gravitational field}

The issue of turbulent combustion in a uniform 
gravitational field 
singles out one important element of supernovae deflagration --
propagation of the flame against gravity. 
The turbulence is not imposed
externally, as in many combustion experiments and theoretical
studies, but is driven internally by the 
buoyancy of burning products,
as in supernovae. This problem can be studied
both by direct numerical simulations and experimentally. Using the 
results of studies of this restricted problem, 
various subgrid numerical models for turbulent combustion can be 
calibrated  and then applied to supernovae.

Consider  a square column of size $L$ filled with a
premixed gaseous fuel in a uniform gravitational field $g$.
The fuel is initially in the state of hydrostatic
equilibrium. After the fuel is ignited at the bottom, 
a flame begins to propagate vertically. 
Because the propagation is against gravity, the flame
becomes turbulent due to the RT instability
 \footnote{A laminar flame front propagating against 
gravity is unstable with respect to perturbations of all
wavelengths exceeding $\simeq 10-100 \delta_l$, 
where $\delta_l$ is the
laminar flame thickness \cite{Landau}\cite{Markstein}.
It is
nonlinearly stabilized due to the difference in the behavior of
concave and convex parts of the 
flame surface and formation of cusps
\cite{Zeld66} for a wavelength of perturbations 
$\lambda < \lambda_c\simeq {6\pi S_l^2\over g} 
{\alpha+1\over\alpha-1}$, where
$S_l$ is the laminar flame speed, $g$ 
is the acceleration of gravity, and 
$\alpha > 1$ is the ratio of densities ahead and 
behind the flame front \cite{Kho95}. 
For supernovae conditions, $\lambda_c \sim 10^4-10^5$ cm $\gg$ $\delta_l$,
depending on density and position of the flame in the star. 
On larger scales,
the flame is truly turbulent. 
The transition to turbulence is facilitated
by the development of secondary
instabilities, and
occurs on a time scale of $\simeq 2-3$  RT characteristic times.
}.

For uniform conditions, the flame would eventually reach a 
steady-state burning velocity, $S_t$, 
which is statistically independent of time. 
The general functional dependence of $S_t$ 
on the parameters of the problem 
follows from dimensional arguments:
$S_t = \sqrt{gL} ~ f(\alpha,{\rm F})$, 
where ${\rm F=S_l^2/gL}$ is the Froude
number which  characterizes the relative importance of
laminar burning at speed $S_l$ and the R-T instability 
to the fluid flow on the largest scales. 
When gravity dominates (${\rm F} \rightarrow 0$), 
the turbulent flame velocity
becomes independent of the 
details of small scale motions, and of the microphysics
(nuclear reaction rates, etc.) of burning. 
The turbulent flame speed becomes a function of $g$, $L$, and
$\alpha$ only,
$$ 
S_t~=~ \sqrt{gL}\, f(\alpha) ~\simeq~
         C \,\left(gL{\alpha-1\over\alpha+1}\right)^{1/2}~.   \eqno(1) 
$$                 
Numerical simulations confirm this 
scaling law and give $C\simeq 0.5$ for a 
square column with periodic boundary 
conditions  \cite{Kho95}\cite{KOW1}.

The mechanism underpinning the scaling law is the
competition of two effects: surface creation by stretch,
and surface destruction
due to self-intersections of the highly distorted flame.  
In a steady state,
these effects balance each other, 
and $S_t$ becomes independent of $S_l$.  The
flame regulates itself so as to create more 
surface if $S_l$ is small, and
creates less surface if $S_l$ is large \cite{Kho95}.
The scaling law suggests a linear 
relation between the turbulent flame speed
and the velocity of turbulent motions on the 
largest scale, $S_t \sim U'(L)$.
It is clear that the self-regulating 
mechanism operates in all situations when
the turbulence is driven on large scales, 
and cascades down to small scales.

\begin{figure}
\vspace{8cm}
\caption{Regimes of combustion on the $f^2/F - L/\delta_l$ plane.
Line (1) -- $f^2/F = 25$. Line (2) -- $\lambda_G = 2 \delta_l$.
Laminar and weakly wrinkled flamelet regimes are below the line (1).
The multiple flame sheet or corrugated flamelet regime is between the lines
(1) and (2). The scaling analysis of \S 2 applies to this regime.
The distributed flame regime is above 
line (2). The possibility of initiating a detonation
inside an active turbulent flame front (\S 3) exists in this regime.
Circle -- conditions of the terrestrial experiment discussed in \S 2.
Triangle -- conditions for turbulent burning in the middle of a supernova with
$\rho=10^8$ g/cm$^3$, at the scale $L=2\times 10^6$ cm.
}
\end{figure}      

For this self-regulation to
take place, the turbulent scale where the turbulent velocity
equals the laminar flame speed, 
also called the Gibson scale \cite{Peters}, can be expressed as:
$$ 
\lambda_G ~\simeq~ L \left( S_l \over S_t \right)^3    ~, \eqno (2) 
$$
must be larger than the laminar flame thickness $\delta_l$.  
This self-regulated regime can be classified as a corrugated
flame regime \cite{Peters}, or as
a regime of multiple flame sheets \cite{Williams}. 

One example of a terrestrial system  well suited for experimental
study of buoyancy-driven turbulent burning is a lean methane flame 
diluted to an equivalence ratio 
$\phi \simeq 0.55 - 0.6$. The laminar flame velocity is 
$S_l \simeq 10$ cm s$^{-1}$,
and the flame thickness is $\delta_l \simeq 1$ mm \cite{Faeth}.
The regime of the scaling
law of equation (1) can be achieved in a vertical tube with $L\simeq 30-50$ cm,
and height of several meters \cite{KOW1}.
This regime is shown in Figure~1 which shows the plane spanned
by the parameters $f^2/{\rm F}$ and $L/\delta_l$. The circle
shows the condition predicted for the methane experiments 
and the triangle shows  burning on the
scale $L=10^6$ cm in the
middle of a supernova ($g=1.9\times 10^9$ cm/s, $S_l=10^6$ cm/s).
Experiments to check the predictions of the scaling law are in progress.

\section{Turbulent combustion in an expanding star}

The uniform gravitational
field approximation could be valid  on spatial scales less then the 
mean radius of the flame $R_f$, and on time scales less than the 
time scale of the expansion $\tau_e\simeq dt/d\ln R_*$, where $R_*$ is
the stellar radius. The main effect of expansion is to freeze turbulence
on the scales where the turbulent velocity generated by the RT instability
is comparable to or less than the expansion velocity on those scales.
We expect this freeze-out to occur on a scale $L_f$ where the turbulent 
burning velocity given by equation (1) becomes comparable with the expansion
velocity on this scale \cite{Kho95}, i.e, where
$$
       S_t(L_f) \simeq U_e(L_f)\equiv L_f / \tau_e~.                   \eqno(3)
$$
To get an estimate of $L_f$ we use for the expansion rate 
the results of a one-dimensional
simulation with the average turbulent
flame speed given by equation (1) with $L$ equal to the flame radius
(assuming no turbulence freeze-out or effects of spherical geometry)
 \cite{Kho95}. This gives $L_f \simeq {\rm (a~few)}\times 10^7$ cm, 
less than the mean radius of the flame $R_f$ 
which exceeds (a~few)$\times 10^8$ cm 
during the final stages of deflagration, and clearly shows that 
freeze-out is important.
 Using equation (1), we then
estimate the effective deflagration speed as
$$
{S_t\over a_s} \simeq 3\times 10^{-2}
                    \left(5\times 10^8{\rm cm/s}\over a_s\right)
                    \left(g\over 10^9{\rm cm/s^2}\right)^{1/2}
                    \left(L_f\over 10^8 {\rm cm} \right)^{1/2}
                                                                     \eqno(4)
$$
where $a_s$ is the sound speed in the stellar matter. In conditions typical of
an exploding white dwarf, we do not expect the turbulent burning speed
to exceed a few per cent of the sound speed.  This is not enough to cause a
powerful explosion even in the idealized one-dimensional picture
where the front is spherical, and  all matter passing through the front is
burned.

An additional effect that further limits the rate of deflagration 
is a deviation from the steady-state turbulent burning regime. 
A certain time,
$\simeq \lambda/S_t(\lambda)$,  is  required for a turbulent
 flame to reach a steady-state on the scale $\lambda$.  This time is
larger for larger scales. Scales of the order of $R_f$ might never reach a
steady-state. This can be also expressed in a different way. The thickness
of a steady state turbulent front in a vertical column of size $L$ is
$\simeq 2-3 L$ \cite{Kho95}\cite{KOW1}; but for $L\simeq R_f$, there is not
enough space in a star to fit a steady-state turbulent flame inside the
star.

Some results of a three-dimensional numerical simulation of the entire
white dwarf exploding as a supernova are presented in Figures 2 and 3
(see \cite{Kho95} for details). In these simulations, equation
(1) has been used  for the turbulent flame
velocity on scales not resolved numerically. For regions of the ``average''
flame front not oriented ``upwards'' against gravity this formula 
most probably overestimates the local turbulent flame speed. 
Despite this,  only $\simeq 5$\% of the mass has been burned by
the time the model has expanded and quenched the flame, and
the white dwarf has not become unbound. These results show that burning
on large scales does not reach a steady state.  Big blobs of burned
gas rise and penetrate low density outer layers, whereas unburned
matter flows down and reaches the stellar center. 
The flame was not able to
close the gaps between the blobs and make the flame surface 
quasi-spherical.  Even if this had occurred, however,
the speed of the deflagration would have been exactly as described
by equation (1) with L given by equation (3), and this is not 
enough to unbind the model. 

The model experienced
an almost complete overturn. This has obviously important implications
for nucleosynthesis and may cause an element stratification 
incompatible with observations if composition inhomogeneities are not 
smeared out during the subsequent detonation stage of burning.

The results of 3D modeling indicate that the deflagration
alone is not sufficient to cause an explosion.  To make a powerful explosion, 
the deflagration must somehow make a 
transition to a detonation (delayed detonation
model). This idea is also supported by observations of SN~Ia, events that
are best described in the framework of the delayed and pulsating
delayed detonation scenarios (\cite{Kho91a}, \cite{Kho91b},
\cite{KHM93}).

\begin{figure}
\vspace{5cm}
\caption{Results of a three-dimensional simulation  of a turbulent
deflagration explosion of a $1.4 M_\odot$ carbon-oxygen white dwarf
(\S 3).
The central density (dashed line) and the amount of burned matter (in per cent)
as a function of time elapsed since ignition in the center are shown.}
\end{figure}

\begin{figure}
\vspace{11.5cm}
\caption{The distribution of burned inside a $1.4 M_\odot$
white dwarf resulting from $\simeq 1.5$ s of 
turbulent combustion for  the simulation shown in Figure 2.
 }
\end{figure}
                      
\section{Deflagration to detonation transition (DDT)}

In terrestrial conditions, DDT is often caused by interaction of a
turbulent flame with obstacles or boundaries. DDT in unconfined
conditions is possible, but more difficult to achieve.
Here we apply the theory of DDT in terrestrial unconfined conditions 
proposed in \cite{KOW-2} to a DDT in supernovae. The theory is based on 
two major assumptions: ({\it i}) The detonation arises from a non-uniform
explosion of a region of a fuel with a gradient of induction time
via the Zeldovich gradient mechanism \cite{Zeld-grad} \cite{SWAZER}. 
This region is created by mixing of fresh fuel and hot products of burning,
and is characterized by the oppositely directed gradients of temperature
and composition. There exists a minimum size of the  region capable of 
generating a detonation, $L_i$. In  \cite{KOW-2} we argue on the basis of
numerical simulations and terrestrial experiments that $L_i$ can be 
correlated with the experimental values of the detonation kernel size 
$l_k$ (the minimum length to propagate a steady-state detonation)
as $L_i \simeq (1-4)l_k$ or with the detonation cell size $l_c$ 
(the size of a single detonation ``cell" defined by the loci of
intersecting triple points in the wake of a detonation front) as
$L_i \simeq 10-40 l_c$, or
with the theoretical value of the one-dimensional detonation wave thickness
$\delta_d$ as $L_i \sim 10^3 \delta_l$. The exact coefficients in these
relations depend on the details of reaction kinetics and equation of state.
({\it ii}) A level of turbulence $U_i$ required to mix the products and fuel 
inside a region of size $L_i$ follows from the requirement that the Gibson
scale given by equation (2) be comparable to the laminar flame thickness
$\delta_l$, or
$$
U_i = K S_l \left(L_i / \delta_l \right)^{1/3}              \eqno(5)
$$
where the coefficient $K\simeq 1-10$ describes the ability of the flame
to stabilize itself against turbulent motions on scales $\simeq \delta_l$.
At this level of turbulence, a sharp interface -- a laminar flame front --
between the fuel and products cannot survive turbulent stretch. The flame front
can be locally broken and extinguished, and the mixing of fuel and
products can occur on scales $>> \delta_l$.

The one-dimensional thickness of a detonation front in a carbon-oxygen fuel 
(C+C reaction zone)
ranges from $\simeq 10^{-2}$ cm at densities $\simeq 10^9$ g/cm$^3$ to $\simeq
10$ cm at $\simeq 10^7$ g/cm$^3$ \cite{Kho89}. 
From this we can roughly estimate $L_i \simeq 10 - 10^4$ cm for
the same density range. The
two-dimensional structure of a detonation front has been
computed for the density $3\times 10^7$ g/cm$^3$ in \cite{Boisseau}. The 
detonation cell size was found to be $\simeq 10$ cm, which gives a slightly
smaller value of $L_i \simeq 10^2$ cm than the one-dimensional estimate
$L_i \simeq 10^3$ cm. It is, however, clear that the estimates are correct 
within an order of magnitude. We also notice the ratio of thicknesses of the
one-dimensional detonation and a laminar flame in a degenerate CO mixture is
approximately $\delta_l/\delta_d \simeq D/S_l \simeq 10^{-2}$.

With these data in mind, we can rewrite equation (5) in numbers as
$$
{U_i\over a_s} \simeq 10^{-2}
                    \left(5\times 10^8{\rm cm/s}\over a_s\right)
                    \left(S_l\over 10^4{\rm cm/s}\right)
                                                         \eqno(6)
$$

The laminar flame speed ranges from $10^7$ to $10^4$ cm/s for densities
$10^9 - 10^7$ g/cm$^3$, respectively \cite{Timmes}. Comparing equations
(4) and (6) we see that it might be possible to
trigger a detonation inside an active turbulent flame front when
burning reaches densities $\simeq 10^7$ g/cm$^3$.

If an active turbulent deflagration fails 
to make the transition to a detonation, stellar
expansion will eventually quench nuclear reactions at densities less than 
$\simeq 10^6$ g/cm$^3$. The star will experience a pulsation and collapse
back. During the expansion and contraction phases of the pulsation, the
high entropy ashes of the 
dead deflagration front will mix with the the fresh low
entropy fuel again to form a mixture with temperature and
composition gradients. Mixing will be facilitated during the
contraction phase because of the increase of turbulent motions due to
the conservation of angular momentum. 
An estimate of the mixing region formed
during pulsation is $\simeq 10^6 - 10^7$ cm, much larger than $L_i$.
 As soon as this mixture returns to high enough densities
$\simeq 10^7$ and re-ignites, the detonation will be triggered
(\cite{Kho91b},\cite{ArnettLivne94}).

The theoretical conclusion is that  DDT should occur in a carbon-oxygen 
white dwarf
one way (delayed detonation) or another (pulsation delayed detonation)
at densities $\simeq 10^7$ g/cm$^3$, but at present it is difficult to say
which mode is realized more frequently. The estimate of the transition
density at which detonation must be triggered is consistent with the
estimate of the transition density derived from fitting one-dimensional
delayed detonation and pulsation delayed detonation models with observations
of SN~Ia. A recent success of the delayed detonation model is the  
explanation of the correlated photometric and spectroscopic diversity
of SNIa ((\cite{KHM93},\cite{HK96},\cite{HKW},\cite{BK95},
\cite{WHKH95},\cite{Wheeler96}; see also contributions by 
 H\"oflich et al.\ in these proceedings).

\section{Conclusions}

Despite substantial progress in hydrodynamical modeling and 
understanding of SNIa, there are many
important problems yet to address. A few important ones are:

\smallskip\noindent
To determine the exact deflagration velocity inside a carbon-oxygen white dwarf
as a function of initial composition and initial conditions. This requires
high resolution 3D modeling in order to correctly represent the
competition between the turbulent cascade and turbulence freeze-out.

\smallskip\noindent
To understand the mode(s) of deflagration to detonation transition in a 
white dwarf.
This would require both large scale modeling of the pulsation, and small scale
 modeling of mixing processes in active and dead deflagration fronts.

\smallskip\noindent
Both efforts will require new numerical techniques to provide a resolution
of many orders of spatial scales in 3D. One such a technique, an Eulerian Tree
adaptive mesh refinement, has been recently developed in Texas and successfully
applied to the 3D problem of stellar disruption (\cite{ajk}). 
It is being applied to both terrestrial and astrophysical 
combustion problems (in progress).

\acknowledgements
The work described in this paper has been done in the framework of the
Terrestrial and Astrophysical Combustion Project sponsored by the NSF,
NASA, the Texas Advanced Research Program and the Naval Research Laboratory.
We thank P.A. H\"oflich, Robert P. Harkness, Geraint Thomas, Eli Livne,
Norbert Peters, Gerard Faeth, Charles Westbrook, Chung Law, Howard Ross
and Jay Boris for important discussions, useful information,
ongoing contributions, encouragement and support.     
JCW is grateful to the organizers for support at this  
stimulating meeting.

\end{document}